\documentclass[proof]{pasj01}
\usepackage{comment}
\usepackage{graphicx}
\usepackage{caption} 
\usepackage{lscape} 
\usepackage[T1]{fontenc} 
\usepackage{mathrsfs}
\usepackage{subfigure} 
\usepackage{subfig} 
\usepackage{adjustbox} 
\usepackage{tabularx} 
\usepackage{booktabs} 
\usepackage{ulem} 
\usepackage{soul} 
\usepackage{xcolor} 
\usepackage{here}  
\usepackage{afterpage} 

\Received{$\langle$reception date$\rangle$}
\Accepted{$\langle$acception date$\rangle$}
\Published{$\langle$publication date$\rangle$}

\begin{document}

\title{Intrinsic color diversity of nearby type Ia supernovae}

\author{Noriaki \textsc{Arima},\altaffilmark{1,2}$^{*}$ Mamoru \textsc{Doi},\altaffilmark{1,3,4}
Tomoki \textsc{Morokuma},\altaffilmark{1}
and Naohiro \textsc{Takanashi}\altaffilmark{5}}

\altaffiltext{1}{Institute of Astronomy, Graduate School of Science, The University of Tokyo, 2-21-1 Osawa, Mitaka, Tokyo 181-0015, Japan}
\altaffiltext{2}{Department of Astronomy, Graduate School of Science, The University of 
Tokyo, 7-3-1 Hongo, Bunkyo-ku, Tokyo 113-0033, Japan}
\altaffiltext{3}{Research Center for the Early Universe, Graduate School of Science, The University of Tokyo, 7-3-1 Hongo, Bunkyo-ku, Tokyo 113-0033, Japan}
\altaffiltext{4}{Kavli Institute for the Physics and Mathematics of the Universe, The University of Tokyo, 5-1-5 Kashiwanoha, Kashiwa, Chiba 277-8583, Japan}
\altaffiltext{5}{Executive Management Program, The University of Tokyo, 7-3-1 Hongo, Bunkyo-ku, Tokyo 113-8654, Japan}

\email{n\_arima@ioa.s.u-tokyo.ac.jp}


\maketitle

\begin{abstract}
It has been reported that the extinction law for Type Ia Supernovae (SNe Ia) may be different from the one in the Milky Way, but the intrinsic color of SNe Ia and the dust extinction are observationally mixed. In this study, we examine photometric properties of SNe Ia in the nearby universe ($z \lesssim 0.04$) to investigate the SN Ia intrinsic color and the dust extinction. We focus on the Branch spectroscopic classification of 34 SNe Ia and morphological types of host galaxies. We carefully study their distribution of peak colors on the $B-V$, $V-R$ color-color diagram, as well as the color excess and absolute magnitude deviation from the stretch-color relation of the bluest SNe Ia. We find that  SNe Ia which show the reddest color occur in early-type spirals and the trend holds when divided into Branch sub-types. The dust extinction becomes close to the Milky-Way like extinction if we exclude some peculiar red Broad Line (BL) sub-type SNe Ia. Furthermore, two of these red BLs occur in elliptical galaxies, less-dusty environment, 
suggesting intrinsic color diversity in BL sub-type SNe Ia.
\end{abstract}
\section{ Introduction }
Type Ia supernovae (SNe Ia) have been used as distance indicators thanks to their high luminosity, which can be observed at cosmological distances ($M_B$ $\approx -19.3$ mag). Luminosity can be standardized after correcting the empirical relation between luminosity and light-curve decline rate: a brighter SN Ia diminishes slowly (so-called Phillips-relation: \cite{Phillips1993,Hamuy1996,Riess1996,Phillips1999}). Their use for cosmological studies provided us an observational evidence that the expansion speed of the Universe is accelerating (Riess et al. 1998; Perlmutter et al. 1999).\par
Near the epoch of maximum brightness, SNe Ia show strong silicon absorption lines, especially the Si {\scriptsize II} $\lambda 6355$ but neither hydrogen nor helium lines are present (\cite{Filippenko1997,Gal-Yam2017}). SNe Ia are thought to originate from carbon-oxygen (C/O) white dwarfs in close binary systems. Theoretically, two most popular SN Ia scenarios
are the single-degenerate (SD) model and the double-degenerate (DD) model. In the SD model, a white dwarf accretes material from a non-degenerate companion star until it reaches near the Chandrasekhar limiting mass ($M_{\rm Ch} \approx 1.4 M_{\odot}$) and explodes (\cite{Whelan1973,Nomoto1982}). In the DD model, two white dwarfs merge after losing orbital energy and angular momentum by gravitational waves (\cite{Jr.Iben1984,Webbink1984}). However, we still don't have clear physical understandings of the evolutionary paths to SNe Ia and there are several other potential progenitor models (see reviews of  \cite{Maoz2012,Maeda2016} and also \cite{Livio2018}). In addition, the Phillips-relation lacks complete physical understanding.\par
Identifying the differences and diversity of SNe Ia is important for understanding their progenitor scenarios and improving their effectiveness as a cosmological tool. Recent cosmological measurements with SNe Ia are not only limited by the sample size, but also by the systematic uncertainties due to the lack of our understanding the variety and physical mechanisms of SNe Ia (\cite{Conley2010,Betoule2014}). Although observables that characterize each SN Ia such as color may give us a clue, scatters in colors and dust properties surrounding SNe Ia prevent us to reduce uncertainties for precision cosmology (e.g., see figure 1 of \cite{Sullivan2011}).\par
It is also critical for SN Ia studies to understand the properties of interstellar and circumstellar dust. In general, extinction by dust is described by a parameter $R_V$: total-to-selective extinction ratio defined by $R_V = A_V/E(B-V)$, which reflects the property of dust. Large value of $R_V$ means large grain size of dust (c.f., \cite{Clayton2003}). The mean value in our Milky Way is $R_V = 3.1$ (\cite{Cardelli1989}). It has been reported by previous studies that the dust extinction for SNe Ia has smaller value of $R_V$ than the mean value of Milky Way. For example, \citet{Nobili2008} and \citet{Kessler2009} reported $R_V = 1.75$ and $R_V = 2.2$, respectively (see also table 1 of \citet{Cikota2016} for the summary of small $R_V$ reported until then). Assuming a constant $R_V$, \citet{Folatelli2010} analyzed the optical–NIR colors of the nearby Carnegie Supernova Project (CSP: \cite{Contreras2010}) SN Ia sample and they obtained $R_V \approx 1.7$, but they obtained $R_V = 3.2 \pm 0.4$ when the extremely red objects $(E(B-V) \gtrsim 1)$ were excluded. \citet{Mandel2011} later did a more sophisticated analysis and showed variations in $R_V$, with the low-extinction events giving higher values of $R_V \sim 3$ and high-extinction events giving $R_V \sim 2$. \citet{Burns2014} found a similar result. In recent studies, \citet{Stanishev2018} used optical-NIR light curves to derive $R_V \simeq 1.8-1.9$ and \citet{Cikota2016} also favors small $R_V$ but they obtained different $R_V$ values for SNe Ia with different host morphology ($R_V = 2.71\pm1.58$ for SNe Ia observed in Sab–Sbp galaxies, and $R_V = 1.70\pm0.38$ for SNe Ia observed in Sbc–Scp galaxies). On the other hand, based on a spectral series, \citet{Sasdelli2016} showed $R_V$ to be consistent with the typical Milky Way value. \citet{Mandel2017} constructed host galaxy dust models for SNe Ia and the dust extinction they obtained ($R_B = 3.8 \pm 0.3; R_B = R_V +1$) also agrees with the Milky Way dust extinction.\par
Over the years, observations provided that there are varieties in luminosity and spectral features of SNe Ia. For example, 1991T-like and 1991bg-like SNe are well-known prominent outliers. The 1991T-like SNe are more luminous ($\gtrsim$ 0.6 mag.)  than normal SNe Ia and show lines of doubly ionized elements (especially Fe {\scriptsize III}) in their spectra at early times (\cite{Filippenko1992, Phillips1992}). As opposed to it, the 1991bg-like SNe are less-luminous and show rapidly evolving light-curve and strong Ti {\scriptsize II} lines (\cite{PeterNugent1995,Mazzali1997}). Diagnosing spectra around maximum brightness has been used to investigate the origins of the diversity.\par
In \citet{Branch2006}, SNe Ia were assigned into four groups according to measurements of the equivalent width (EW) of two Si  {\scriptsize II} absorption features at about 5750 \AA \  and 6100 \AA \ which are attributed to rest-frame $\lambda 5972$ and $\lambda 6355$ lines, respectively. The four groups are: core-normal (CN), broad-line (BL), cool (CL), and shallow-silicon (SS). The 1991T-like and the 1991bg-like SNe Ia are assigned into the extreme end of SS and CL, respectively. Using early phase SN Ia colors, \citet{Stritzinger2018} found that there are two distinct populations with different early color evolution in $B-V$, and the two early blue/red events are correlated with the Branch spectroscopic groups.\par
Another spectroscopic approach is dividing SNe Ia into two groups in terms of the expansion velocity estimated from the absorption minimum of Si {\scriptsize II} $\lambda 6355$ line. \citet{Wang2009} found that high velocity (HV) SNe Ia show redder $B-V$ colors at maximum brightness than normal velocity (NV) SNe Ia. \citet{Zheng2018} found that SNe Ia with higher velocities are inferred to be intrinsically fainter than the NV SNe Ia and they confirm that HV SNe Ia are probably intrinsically different from NV SNe Ia.\par
Focusing on the differences in environment is another approach to study the diversity of SNe Ia. It is found that host galaxy stellar mass correlates with SNe Ia luminosity: SNe Ia in massive host galaxies are intrinsically more luminous after light-curve/color corrections (\cite{Kelly2010,Sullivan2010,Childress2013,Pan2014,Betoule2014}). The latest cosmological study by \citet{Smith2020} found that SNe Ia in high-mass galaxies ($>10^{10}\ M_{\odot}$) are intrinsically more luminous than their low-mass counterparts by $0.040\pm 0.019$ mag. Correlations with host-galaxy metallicity have also been studied (e.g., \cite{Moreno-Raya2016}). Theoretically, the metallicity of the SN Ia progenitors affects the strength of spectral features more in the UV wavelengths than in the  optical (Lentz et al. 2000; Walker et al. 2012; Miles et al. 2016). However, each model has a different effect on the strength and wavelength range. A recent study using SNe Ia with normal light-curve shapes found that there is no significant correlations between the UV-optical colors of SNe Ia and the host-galaxy metallicity (\cite{Brown2020}), which is in contrast to the findings of \citet{Pan2020}. The physical relation between host metallicity and SN Ia properties is not yet clear.\par
For fitting a light-curve shape to standardize SN Ia luminosity, "MLCS2k2" (\cite{Jha2007}), "SALT2" (\cite{Guy2007}) and "SNooPy" (\cite{Burns2011,Burns2014}) are methods often used. Each method parameterizes observed light curves and estimates host-galaxy dust extinction at the same time. When we investigate the host-galaxy dust extinction, we should treat dust extinction effect and SN Ia intrinsic colors independently, and hence we should not assume any dust extinction models. In contrast, \citet{Takanashi2017} (hereafter TAK17) simply parameterizes only light curve shapes and peak brightness. They analyzed multi-band light curves of SNe Ia from SDSS-II SN Survey and their result suggests that there seems to be inherently different sub-types among SNe Ia with different colors and extinction laws of host galaxy dust, which is consistent with previously suggested ideas (e.g., \cite{Mannucci2005,Quimby2007}).\par
As we mentioned above, the diversity of SN Ia spectra, luminosity and colors may be correlated with host properties and also different types of dust extinction. In order to improve accuracy as distance indicators for cosmological studies, understanding the diversity is very important. In TAK17, they only studied photometric properties. In this paper we extend the study of TAK17 combining with the spectroscopic classification of SNe Ia defined by \citet{Branch2009}. In Section 2, we describe our SNe Ia sample with host galaxy morphology used in this study. In Section 3, we apply the method used in TAK17 to analyze photometric SNe Ia data and show the results of nearby SNe Ia with spectral classification and host morphology. In Section 4, we discuss our results and we summarize our findings in Section 5.
\section{ Data }
\subsection{ Photometric data }
We use the photometric data obtained from \cite{Takanashi2008} (hereafter, TAK08). In TAK08, they collected $U-,B-,V-,R-$ and $I$-band photometry of 122 nearby ($z < 0.11$) SNe Ia from published sources. Magnitudes are presented in the Vega system, and are all $K-$corrected. As we mentioned in Section 1, well-known light curve fitting methods fit observed SNe Ia light curves by parameterizing their light-curve shape, peak brightness, color and extinction at the same time.
In TAK08 (and also in TAK17), unlike these methods, their original "Multi-band Stretch Method" simply characterizes light-curve shapes and peak brightness corrected for Galactic dust extinction from \citet{Schlegel1998} without dust extinction correction in the host galaxies. This makes it possible to investigate the diversity of SN Ia light curves directly.\par
We use $B-,V-$ and $R-$band absolute magnitude and $B-$band stretch factor (\cite{Goldhaber2001}), which is a parameter describing a width of SN Ia light-curve shape. In TAK08, they adopt the values of cosmological parameters, $H_0 = 70.8\ {\rm km\ s^{-1}\ Mpc^{-1}} $, ${\rm \Omega_M} = 0.262$, ${\rm \Omega_{\Lambda}} = 0.738$ from \citet{Spergel2007} for calculating absolute magnitude. We adopt these values in this paper.
\subsection{ Spectroscopic data }
As described in section 1, \citet{Branch2006} divide SNe Ia into four groups based on equivalent widths (=EWs) of the two Si {\scriptsize II} $\lambda 6355,\lambda 5972$ lines (= EW(6100) and EW(5750)); Core Normal (CN), Broad Line (BL), Cool (CL) and Shallow Silicon (SS). In this paper, we use the term "sub-type" to refer to each Branch spectroscopic group. CN SNe Ia have typical EWs for both Si {\scriptsize II} lines. BL SNe Ia have 6100 \AA \ absorption that is broader and deeper than CN SNe Ia and that means they have large EW(6100). For CL SNe Ia, the name "Cool" comes from their low temperature compared with other sub-types and they show relatively large EWs for both lines. On the other hand, SS SNe Ia have shallow Si {\scriptsize II} absorption lines, which implies the ejecta is in high temperature. In \citet{Branch2009}, they say that there is no distinct threshold of these two equivalent widths that classifies them into four sub-types.\par
In this study, combining photometry from TAK08 with the Branch spectroscopic sub-types (\cite{Branch2009}), we collect nearby ($0.0021 < z < 0.031$) 34 SNe Ia (hereafter, referred to as "Branch sample"). The Branch sample consists of 9 CN, 10 BL, 3 CL and 12 SS SNe Ia. Although there are slightly different spectroscopic classifications among \citet{Branch2009} and the other papers, we adopt the classification presented in \citet{Branch2009} to keep consistency. In \citet{Branch2009}, most of the spectra taken $\pm3$ days from maximum phase are used\footnote{Some of the Branch samples (SN 1997br, 1997do, 1998ab, 1999gh, 2000cn and 2001V) have no Si {\scriptsize II} EWs measurements within $\pm3$ days from maximum phase.}, while for expamle, in \citet{Blondin2012}, spectra taken within $\pm5$ days from maximum phase are used. The properties of our sample are summarized in table 1.
\subsection{ $\Delta M_B$ and $SNCE$ }
In TAK17, they analyzed multi-band light curves of 328 SNe Ia observed by the Sloan Digital Sky Survey-II Supernova Survey (\cite{Sako2008,Sako2018}). Equations (1) and (2) show the relations between the inverse $B-$band stretch factor $s_{(B)}^{-1}$ and $B-$, $V-$band absolute magnitude of the SDSS bluest SNe Ia, which they call stretch-magnitude relations. The color range of the bluest SNe Ia is $-0.14 \leq (M_B - M_V) \leq -0.10$. They introduced two parameters: $\Delta M_B$ and $SNCE$ (= SuperNova Color Excess). They define $\Delta M_B$ as the magnitude difference between $B-$band absolute magnitude, $M_B$ and estimated absolute magnitude with the stretch-magnitude relation of the bluest SN Ia sample. Also, $SNCE$ is defined by the difference between observed $B-V$ color and intrinsic color from the stretch-color relation.
The idea is based on an assumption that relations among a light-curve shape parameter, color and luminosity obey a unique relation for all SNe Ia. The relation can be derived from the bluest SNe Ia, for which extinction may be negligible. In case of the bluest SNe Ia, by definition, both $\Delta M_B$ and $SNCE$ become close to zero. Non-zero $\Delta M_B$ and $SNCE$ can be interpreted as either dust extinction or intrinsically different luminosity and color from the bluest SNe Ia.
\begin{eqnarray}
  M_{B(\rm{bluest\ SNe\ Ia})} &=& (1.71 \pm 0.31) \times s^{-1}_{(B)} - (20.73 \pm 0.31) \\
  M_{V(\rm{bluest\ SNe\ Ia})} &=& (1.59 \pm 0.26) \times s^{-1}_{(B)} - (20.51 \pm 0.25)
\end{eqnarray}
Using the equations (1) and (2), the $\Delta M_B$ and the $SNCE$ are derived as equations (3) and (4)
\begin{eqnarray}
  \Delta M_B &=& m_B - \mu - (1.71 \times s^{-1}_{(B)} - 20.73) \\
  SNCE &=& (m_B - m_V) - (0.12 \times s^{-1}_{(B)} - 0.22)
\end{eqnarray}
, where $m_B$ and $m_V$ are stretch-corrected apparent $B-$band and $V-$band magnitudes at $B-$band maximum and $\mu$ is distance modulus. We derive the $\Delta M_B$ and $SNCE$ using the relations. The list of $\Delta M_B$ and $SNCE$ of the Branch sample is shown in table 2.
\subsection{ Host galaxy morphology }
Because there are strong correlations between SNe Ia and their host galaxy properties (e.g., Sullivan et al. 2006), the inclusion of host galaxy morphology will give us implications to study the SN Ia intrinsic color diversity. We obtained the Hubble type of host galaxies from the NED (NASA/IPAC Extragalactic Database) and then use their host type index $T$ (\cite{De_Vaucouleurs_1959}) to divide SNe Ia into different host morphological groups. We divide SNe Ia into three groups: early-type galaxies (E/S0; $T \leq 0$), early-type spirals (from Sa to Sb; $1 \leq T \leq 4$) and late-type spirals or irregular galaxies (Sc or later and Irr; $T \geq 5$).
\section{ Results }
\subsection{ Color-color diagram }
\begin{figure}[h]
    \centering
	\includegraphics[width=\linewidth]{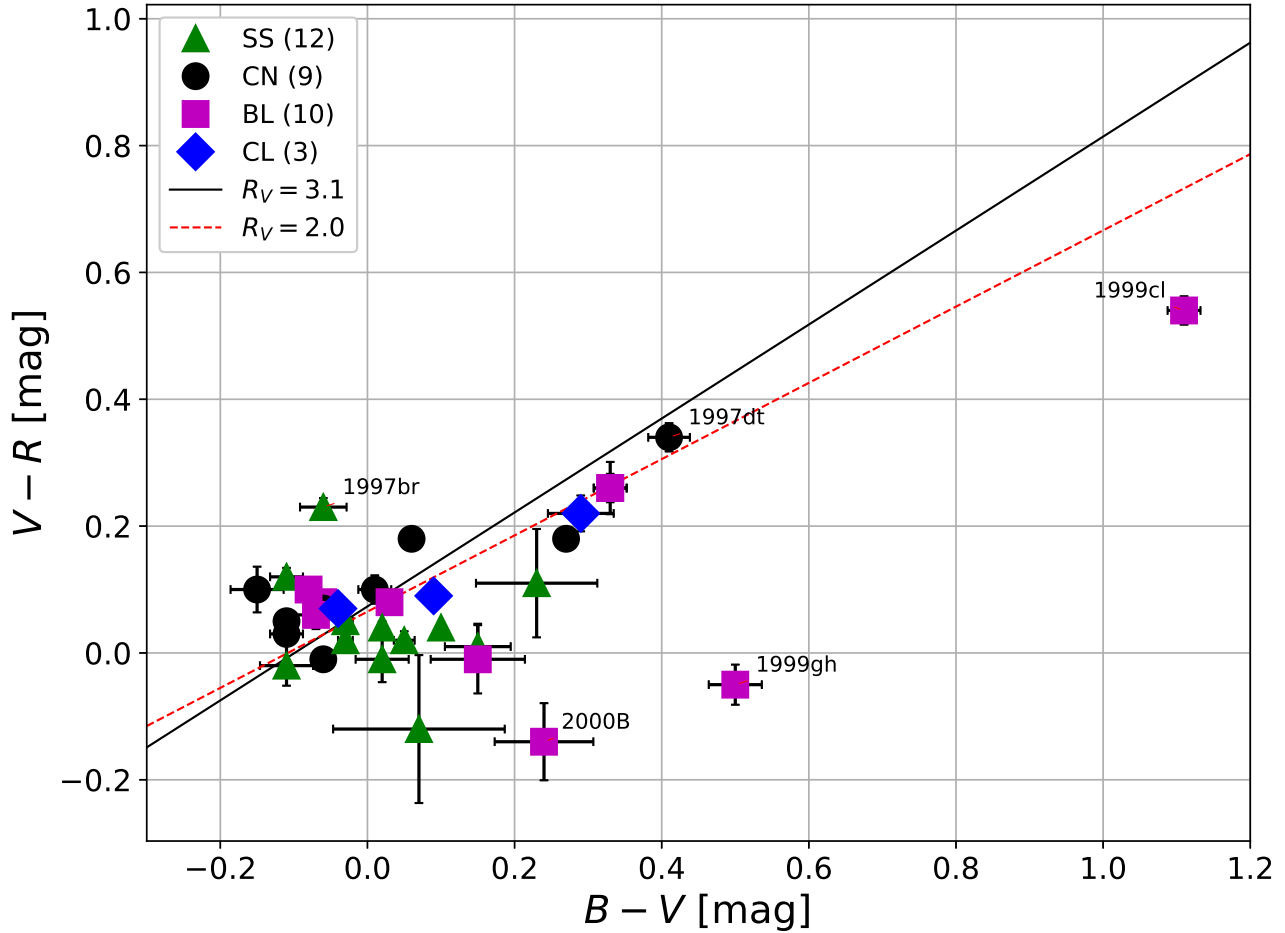}
    \caption{$B-V$ vs $V-R$ color-color diagram. Different symbols show different Branch sub-types. Numbers in the brackets indicate the number of each Branch sub-type sample. Black solid and red dashed lines represent dust extinction laws with $R_V = 3.1$ (Milky-Way like) and $R_V = 2.0$ calculated using \citet{Cardelli1989}, respectively. We draw these lines in order to pass across the main cluster around $B-V \sim V-R \sim 0$.}
    \label{fig:cc_plot}
\end{figure}

In figure \ref{fig:cc_plot}, we show a $B-V$ vs $V-R$ color-color diagram. Two different extinction laws with $R_V  = 3.1$ and $R_V = 2.0$ are shown by black solid and red dashed lines respectively. Extinction law with larger $R_V$ has steeper slope in this diagram. In figure \ref{fig:cc_plot}, the distribution seems to be more consistent with the extinction law with $R_V = 2.0$ than $R_V = 3.1$ for our sample. There are, however, apparently three BL objects (SN 1999cl, 1999gh and 2000B) that are out of the "main sequence" of the distribution. These three BLs are on the lower side of $R_V=2.0$ line and this result is consistent with previous studies that SNe Ia with high Si {\scriptsize II} $\lambda 6355$ expansion velocity have redder colors and prefer a lower $R_V$ (e.g., \cite{Wang2009,Foley2011}). Note that though relatively large EW(6100) labels SN 1999gh as BL, SN 1999gh also has a large EW(5750), which moves it up into the CL sub-type region in the Branch diagram (see figure 2 of \cite{Branch2009}). SN 2000B has no EW measurements in \citet{Branch2009}. CL objects have intrinsically red colors because of their low temperature. If SN 1999gh and 2000B are actually CL sub-type, then the distribution is attributed mostly to intrinsic color rather than interstellar dust.\par
\begin{figure}[h]
    \centering
	\includegraphics[width=\linewidth]{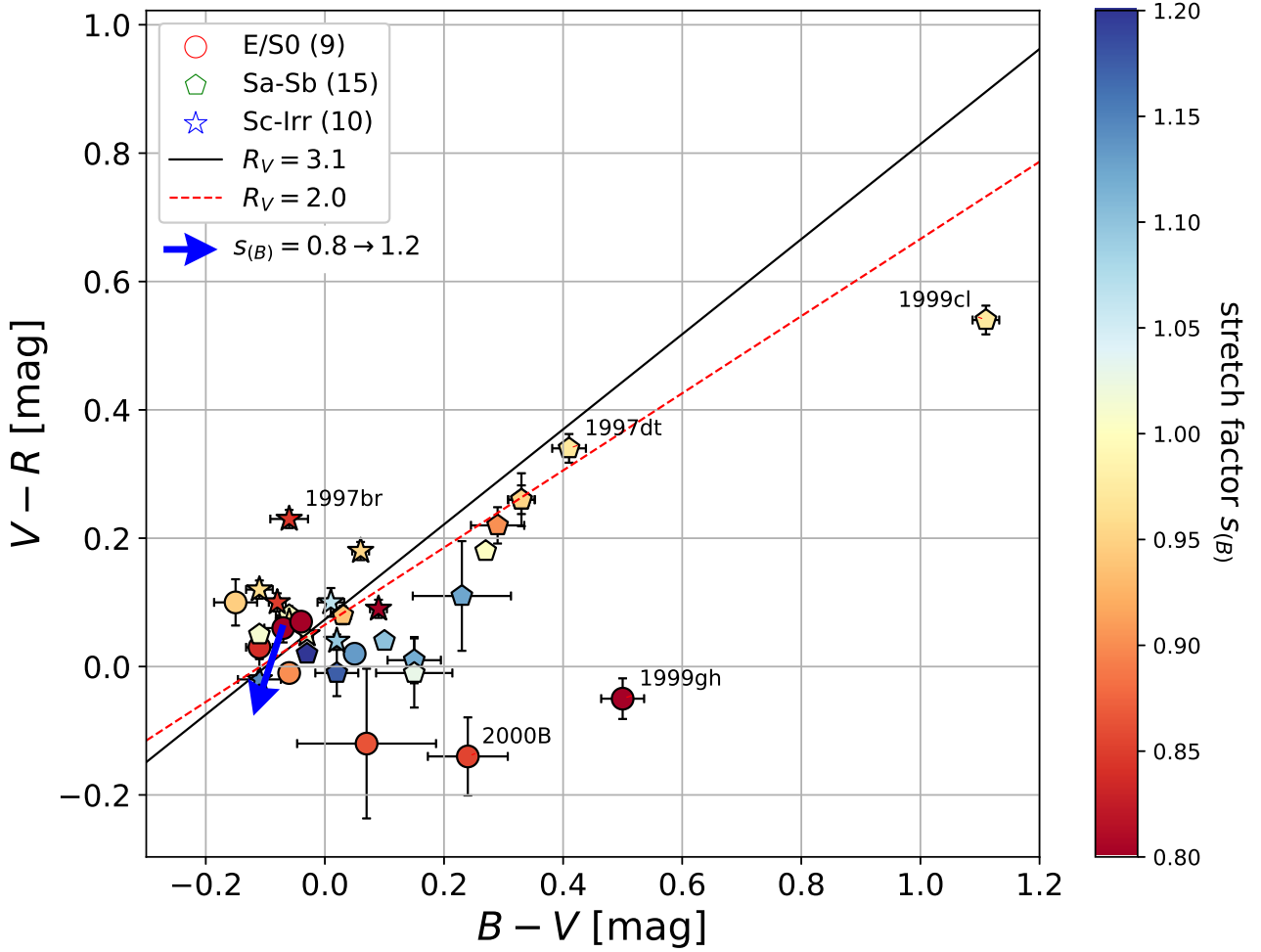}
    \caption{$B-V$ vs $V-R$ color-color diagram with host galaxy morphology and $s_{(B)}$. Circles, pentagons and stars indicate objects whose hosts are elliptical or lenticular galaxies (E/S0; $T \leq 0$), early-type spirals (from Sa to Sb; $1 \leq T \leq 4$), and late-type spirals or irregular galaxies (Sc or later and Irr; $T \geq 5$), respectively. The color bar indicates the $B-$band stretch factor $s_{(B)}$ of SNe Ia. For reference, the blue arrow shows the variation of the bluest SNe Ia colors when $s_{(B)}$ varies from 0.8 to 1.2. The lines presented here are the same as in figure \ref{fig:cc_plot}.}
    \label{fig:cc_host_stretch_plot}
\end{figure}
\begin{figure*}[h] \centering
    \begin{subfigure}[]
        \centering
        \includegraphics[clip,width=0.45\columnwidth]{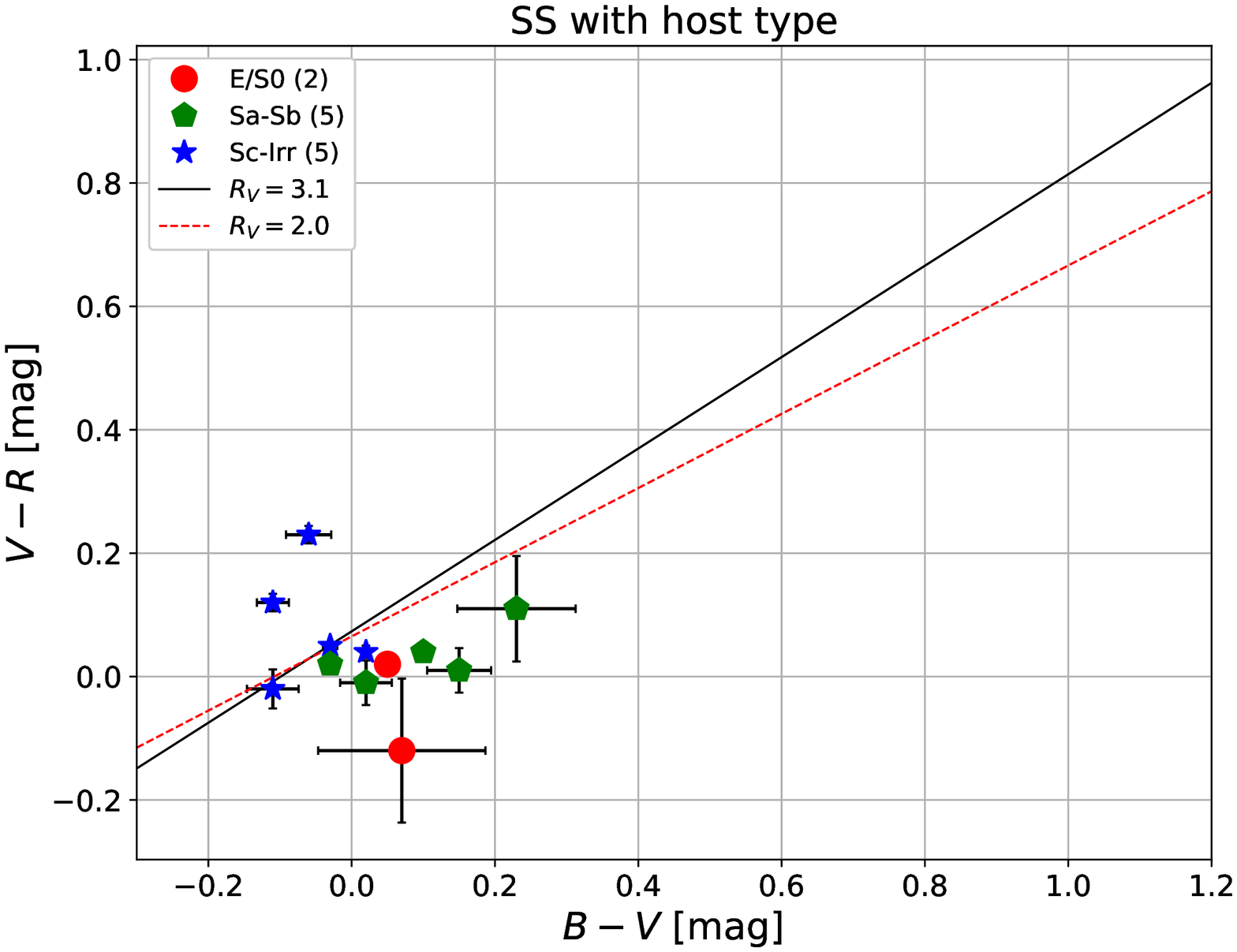}%
    \end{subfigure}
    \begin{subfigure}[]
        \centering
        \includegraphics[clip,width=0.45\columnwidth]{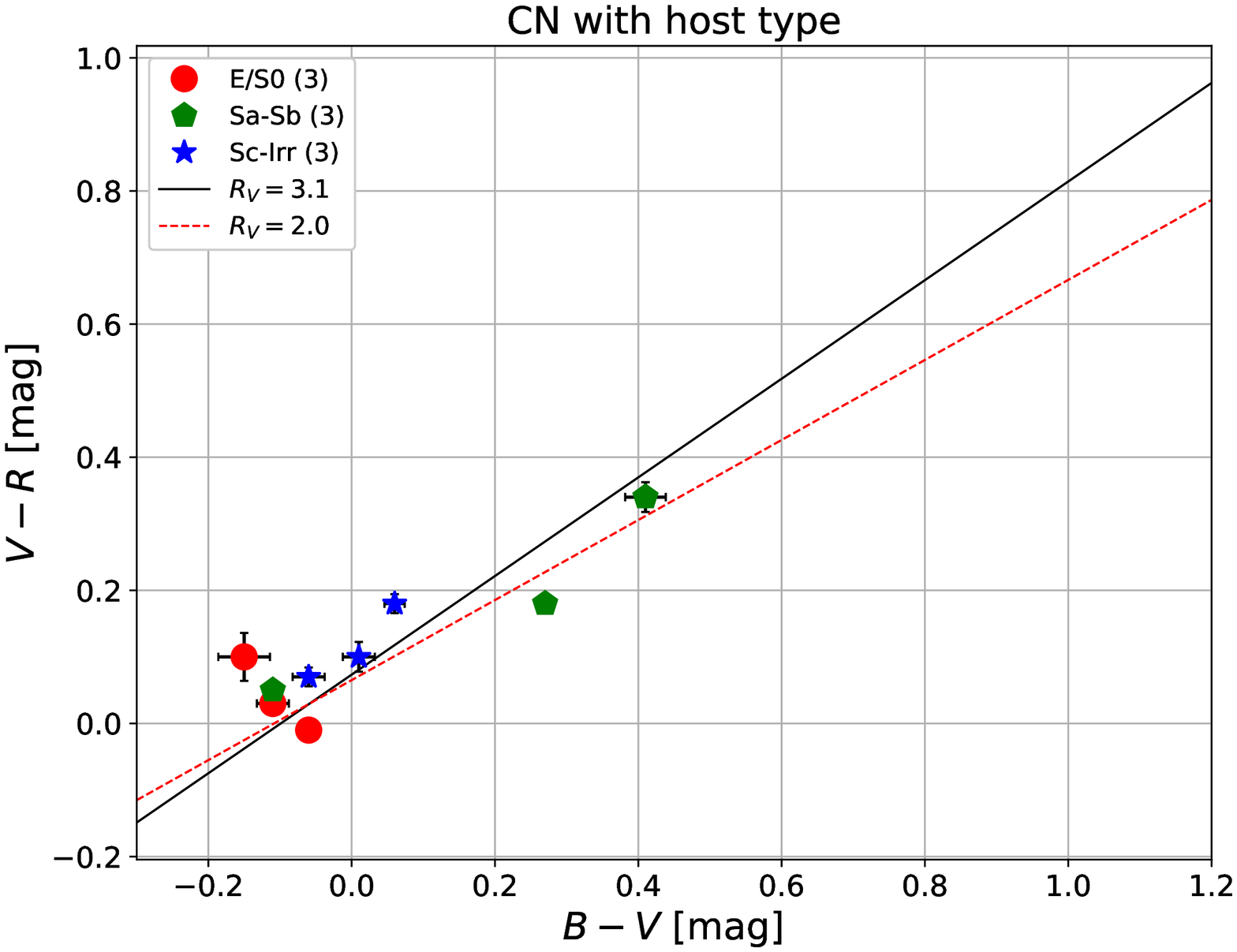}%
    \end{subfigure}
    \begin{subfigure}[]
        \centering
        \includegraphics[clip,width=0.45\columnwidth]{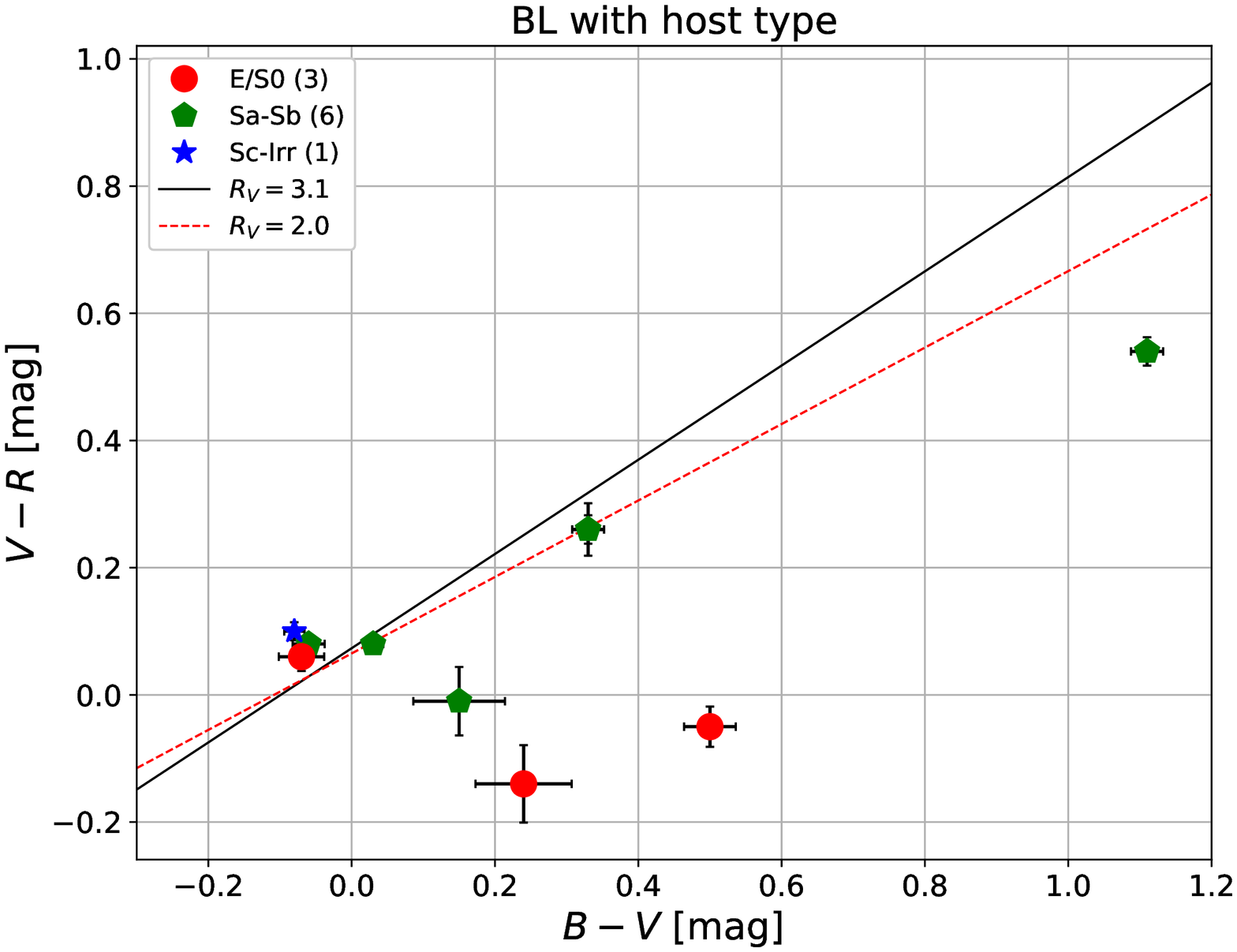}%
    \end{subfigure}
    \begin{subfigure}[]
        \centering
        \includegraphics[clip,width=0.45\columnwidth]{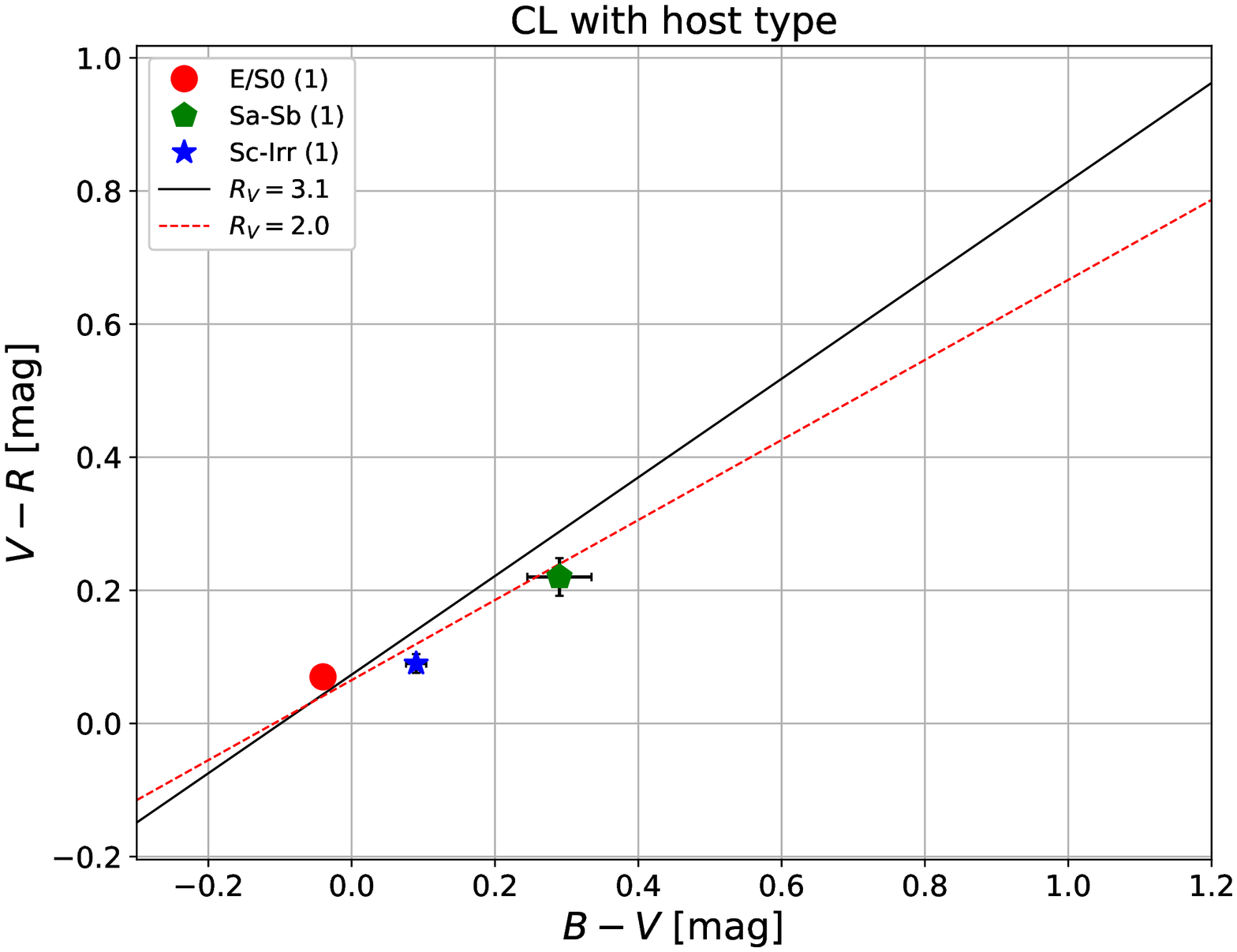}%
    \end{subfigure}
    \caption{$B-V$ vs $V-R$ diagrams with host galaxy morphology. Each figure shows the Branch sample of (a) Shallow Silicon, (b) Core Normal, (c) Broad Line and (d) Cool objects, respectively.  The red and black lines presented here are the same as in figure \ref{fig:cc_plot}.}
    \label{fig:cc_host_plot}
\end{figure*}

Next we examine the color dependence on host morphology. In figure \ref{fig:cc_host_stretch_plot}, different symbols show the different host morphological groups (circles; E/S0, pentagons; from Sa to Sb and stars; Sc or later and Irr). A trend can be seen that SNe Ia whose hosts are early-type spirals (pentagons) have the reddest color along with the dust extinction with $R_V = 2.0$. This result implies that there is intrinsic color offset depending on the host galaxy morphology as it has been said in previous studies (e.g., \cite{Sullivan2010,Pan2014}). The breakdown of the host morphological dependence is shown in figure \ref{fig:cc_host_plot}. Although the sample size is not large, figure \ref{fig:cc_host_plot} shows the trend seen in figure \ref{fig:cc_host_stretch_plot}, SNe Ia which occur in early-type spirals have the reddest color, holds even after divided into Branch sub-types.\par

The effect of $s_{(B)}$ on the $B-V$ and $V-R$ colors is inferred from equations (1), (2) and the $R-$band regression of the bluest SNe Ia\footnote{$(M_B - M_V) = 0.12 \times s_{(B)}^{-1} - 0.22$ and $(M_V - M_R) = 0.34 \times s_{(B)}^{-1 }- 0.36$}. In figure \ref{fig:cc_host_stretch_plot}, the blue arrow shows the variations of colors for the bluest SNe Ia when $s_{(B)}$ varies from $s_{(B)} = 0.8$ to $s_{(B)} = 1.2$. It is expected that colors for most objects will move towards the dust extinction lines. 
The two BL objects (SN 1999gh and 2000B) however, have small $s_{(B)}$ ($s_{(B)} = 0.756$ and $0.854$, respectively) and stretch effect cannot explain the peculiar red colors.\par

\begin{figure}[h]
    \centering
	\includegraphics[width=\linewidth]{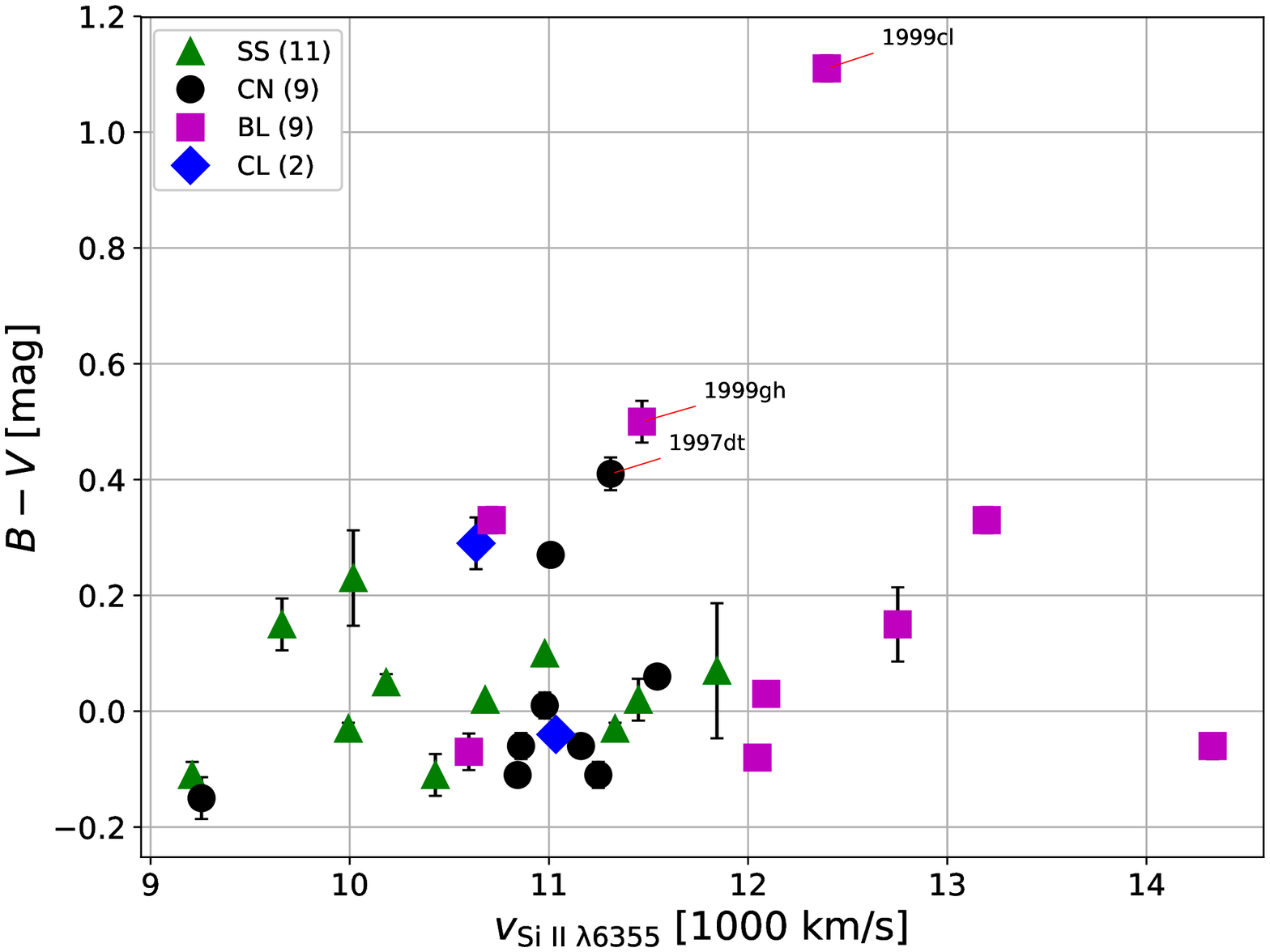}
    \caption{The relation of the Si {\scriptsize II} $\lambda 6355$ absorption line velocity (hereafter $v_{6355}$) and $B-V$ color. The value of $v_{6355}$ is derived from \citet{Blondin2012}. SN 1997br, 2000B and 2000cn have no velocity measurements.}
    \label{fig:Si_velocity}
\end{figure}

We also investigate the relation between $B-V$ color and Si {\scriptsize II} $\lambda 6355$ absorption line velocity. We use the Si {\scriptsize II} velocity measured within $\pm$7 days from maximum brightness phase obtained by \citet{Blondin2012} (see also table 1). In figure \ref{fig:Si_velocity}, SN 1999cl and 1999gh BL objects which show peculiar red $B-V$ and $V-R$ colors have typical values of Si {\scriptsize II} velocity (SN 2000B has no velocity measurements).
\subsection{ $\Delta M_B$ vs $SNCE$ }
\begin{figure}[h]
    \centering
	\includegraphics[width=\linewidth]{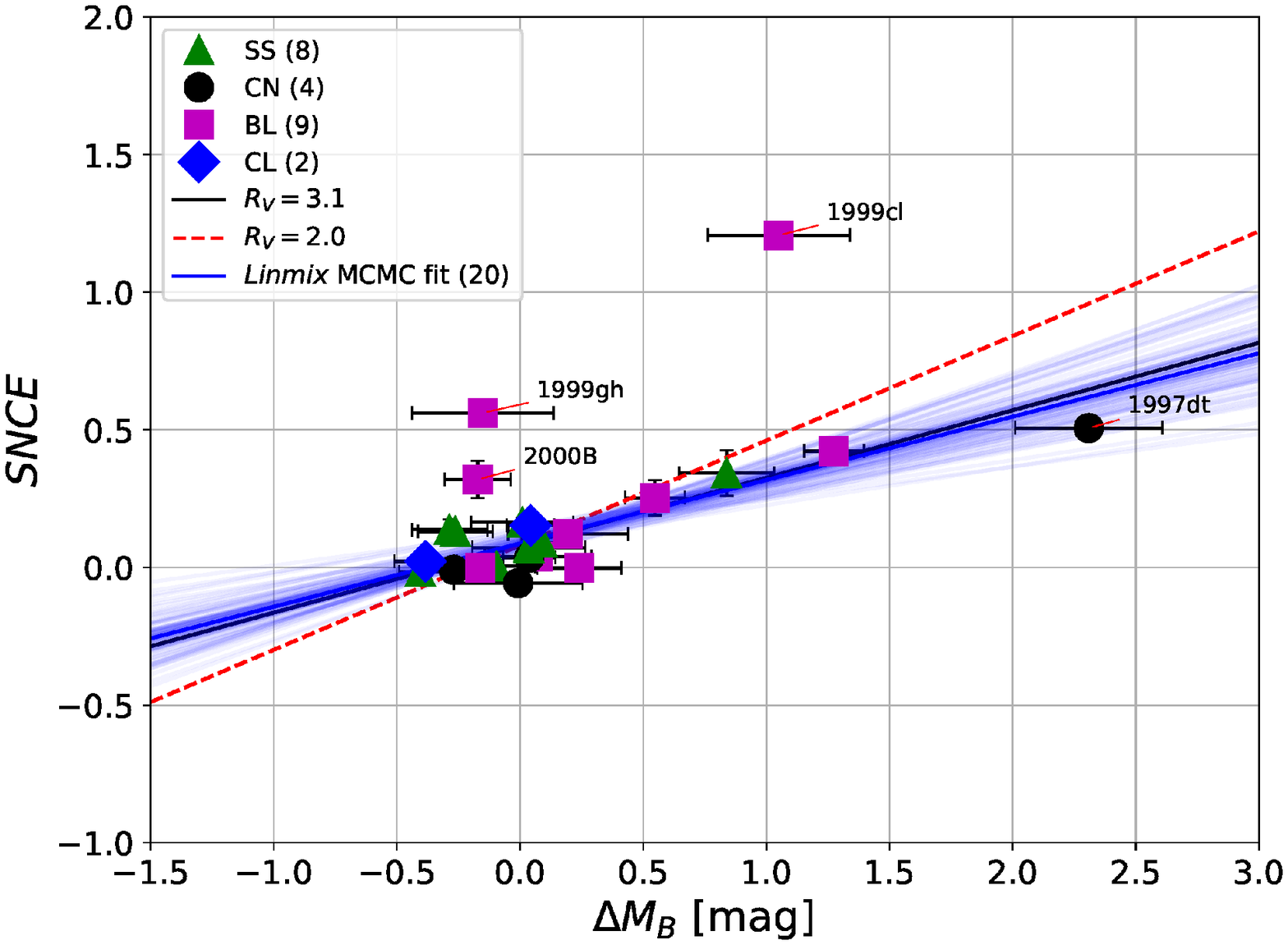}
    \caption{The $\Delta M_B - SNCE$ diagram of the Branch sub-sample. Black solid and red dashed lines denote different dust extinctions with $R_V= 3.1$ and $R_V= 2.0$ respectively. The solid blue line shows the average of linmix MCMC fit results to the 20 samples excluding three BL outliers (SN 1999cl, 1999gh, and 2000B), with 100 random MCMC fit lines overlaid as light blue.}
    \label{fig:dmb_snce_Branch_sample}
\end{figure}

In this section, we examine the $\Delta M_B$ and $SNCE$ parameters introduced by TAK17.
$\Delta M_B$ is sensitive to distance to an object, so nearby SN Ia has a large uncertainty of $\Delta M_B$ since it will be affected by the peculiar velocity ($\sigma_V$) of its host galaxy. Therefore, we plot samples whose total  magnitude uncertainty, $\sigma_{total} = \sqrt{\sigma_{m}^2 + \sigma_{phot.}^2}$ is less than 0.30 mag. $\sigma_m$ is a magnitude uncertainty caused by the peculiar velocity $\sigma_m = \frac{5\ \sigma_V}{cz\ \rm{ln(10)}}$, where $c$ is the speed of light and $z$ is the redshift. $\sigma_{phot.}$ is an uncertainty from photometry. Here we adopt a typical peculiar velocity in the local Universe $\sigma_V = 300\ \rm{km/s}$ and 23 SNe Ia in the Branch sample passed the criterion (hereafter referred to as "Branch sub-sample"). We plot the Branch sub-sample on the $\Delta M_B$ and $SNCE$ diagram in figure \ref{fig:dmb_snce_Branch_sample}.\par
In TAK17, they showed statistically that the sample whose host color is blue (red) has a small (large) value of $SNCE$ and the both samples may obey multiple dust extinction laws. Using SNe Ia with typical stretch factor ($0.9 < s_{(B)} < 1.1$), they obtained extinction law with $R_V = 2.0^{+0.2}_{-0.1}$ from the sample whose host color is red ($u - r > 2.5$) and that with $R_V = 3.7^{+0.3}_{-0.4}$ from the sample whose host color is blue ($u - r < 2.0$).
In figure \ref{fig:dmb_snce_Branch_sample}, we show different dust extinction laws. Black solid line denotes dust extinction similar to that of our Milky-Way ($R_V = 3.1$) and red dashed line shows an extinction law with small $R_V = 2.0$. Most of our samples become to be consistent with the Milky-Way like extinction law after stretch correction. On the other hand, there are some outliers whose $SNCE$ are large (SN 1999cl, 1999gh and 2000B) compared to the other objects with similar $\Delta M_B$.\par
\begin{figure}[h]
    \centering
	\includegraphics[width=\linewidth]{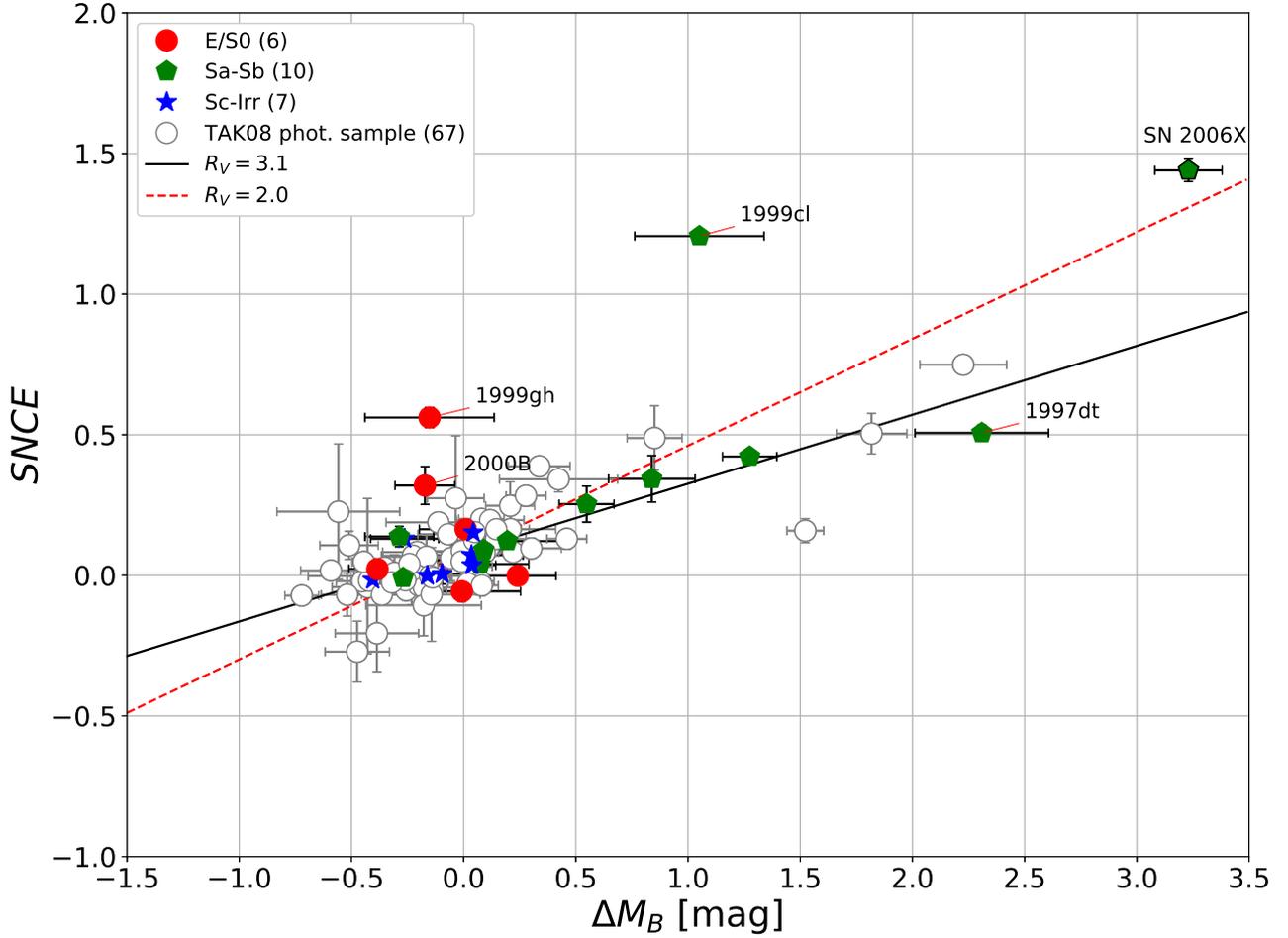}
    \caption{The $\Delta M_B - SNCE$ diagram colored by the host galaxy morphology. TAK08 photometric sub-sample is shown by white filled circles. We also include SN 2006X, a BL sub-type SN Ia with high-extinction for comparison (see section 4.2).}
    \label{fig:dm_snce_host}
\end{figure}

In figure \ref{fig:dm_snce_host}, we plot the $\Delta M_B$ and $SNCE$ diagram colored by the host morphology. The objects whose hosts are early-type spirals (Sa-Sb) are distributed along with the Milky-Way like extinction line. Among the three BL outliers with peculiar red $SNCE$ colors, two objects (SN 1999gh and 2000B) are occured in early-type (E/S0) galaxies.
To increase sample size, we make a subset of 67 objects from TAK08 with $\sigma_{total} < 0.30$ mag (hereafter refereed to as "TAK08 photometric sub-sample") which will be discussed in section 4.3. The list of the TAK08 photometric sub-sample is shown in table 3. We added TAK08 photometric sub-sample shown by white filled circles in figure 6 and 7.\par
\begin{figure}[h]
    \centering
	\includegraphics[width=\linewidth]{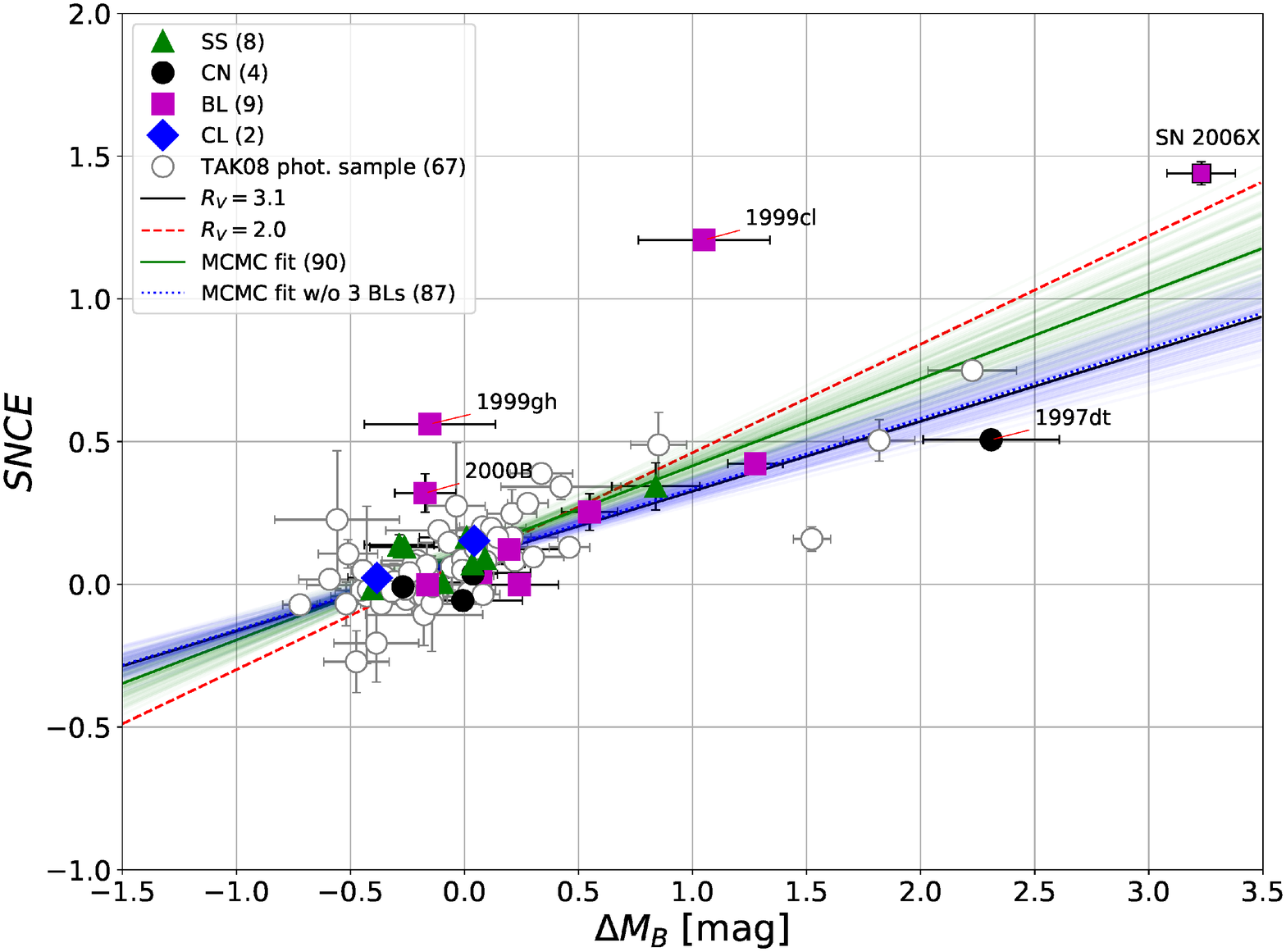}
    \caption{The $\Delta M_B - SNCE$ diagram including the TAK08 photometric sub-sample marked as white filled circles. The green lines show the result of MCMC fit to all Branch sub-sample + TAK08 photometric sub-sample while the blue lines show the result without the three BL outliers. For more detailed descriptions of other lines, see the caption in figure \ref{fig:dmb_snce_Branch_sample}. We also include SN 2006X, a BL sub-type SN Ia with high-extinction for comparison (see section 4.2).}
    \label{fig: dmb_snce_all_sample}
\end{figure}

\subsection{ The distribution of $SNCE$ }
In TAK17, they suggest that there may be different populations on the $\Delta M_B - SNCE$ diagram. That means there may be sub-types with intrinsically different color and dust extinctions. We examined the distribution of $SNCE$ difference between CN and BL sub-types, both of which have typical values of stretch factor $s_{(B)}$. We apply Kolmogorov–Smirnov (KS) test to check if their probability distributions are different or not. As a result, we found the p-value is 0.41 and there is no difference in their distributions at the 5\% significant level. However, our sample is not enough in numbers, so it is desired to obtain the larger number of sample with Branch sub-type classification to test the validity. 
\renewcommand{\arraystretch}{0.7}
\begin{landscape}
\begin{table}[h]
\label{table:our_sample}
\caption{Photometric and spectroscopic information of the Branch sample and the host galaxies}
\scalebox{0.85}{
\begin{tabularx}{1.23\linewidth}{lcccccccccccccc}
\hline
\addlinespace[3pt]
\shortstack {SN \\ name} & \shortstack{Branch\footnotemark[$*$] \\ sub-type} & $s_{(B)}$\footnotemark[$\dag$] & $B$\footnotemark[$\ddag$] & $V$\footnotemark[$\S$] & $R$\footnotemark[$\|$] & \shortstack{EW(6100)\footnotemark[$\#$] \\ (\AA)} & \shortstack{EW(5750)\footnotemark[$**$] \\ (\AA)} & \shortstack{$v_{6355}$\footnotemark[$\dag\dag$] \\ (km/s)} & \shortstack{Phase\footnotemark[$\ddag\ddag$] \\ of spectrum (day)} & \shortstack{Wang\footnotemark[$\S\S$] \\ class} & \shortstack{Host\footnotemark[$\|\|$] \\ name} & \shortstack{$cz$\footnotemark[$\#\#$] \\ (km/s)} & \shortstack{Host\footnotemark[$***$] \\ morphology} &
\shortstack{Host \\ $T$ index} \\ 
\hline
1989B  & CL & 0.902(0.021) & -18.69(0.04) & -18.98(0.02) & -19.20(0.02) & 126 & 25 & -10634 & -1.3 & N   & NGC 3627      & 727  & SAB(s)b                    & 3  \\
1991T  & SS & 1.119(0.025) & -20.77(0.04) & -20.92(0.02) & -20.93(0.03) & 27  & 3  & -9660  & -1.5 & 91T & NGC 4527      & 1736 & SAB(s)bc                   & 4  \\
1994D  & CN & 0.838(0.010) & -18.49(0.02) & -18.38(0.01) & -18.41(0.01) & 98  & 21 & -11248 & -0.1 & N   & NGC 4526      & 617  & SAB(s)0\textasciicircum{}0 & -2 \\
1994ae & CN & 1.067(0.015) & -18.68(0.01) & -18.69(0.02) & -18.79(0.01) & 87  & 10 & -10979 & 0.0  & N   & NGC 3370      & 1279 & SA(s)c                     & 5  \\
1996X  & CN & 0.900(0.006) & -19.63(0.01) & -19.57(0.01) & -19.56(0.01) & 85  & 20 & -11161 & 0.1  & N   & NGC 5061      & 2082 & E0                         & -5 \\
1997br & SS & 0.845(0.016) & -19.37(0.03) & -19.31(0.01) & -19.54(0.01) & -   & -  & -      & -    & -   & ESO 576-40    & 2085 & Sd/Irr                     & 7  \\
1997do & BL & 1.002(0.011) & -18.95(0.02) & -18.89(0.01) & -18.97(0.01) & -   & -  & -14332 & -6.6 & HV  & UGC 3845      & 3034 & Sbc                        & 4  \\
1997dt & CN & 0.971(0.009) & -16.66(0.02) & -17.07(0.02) & -17.41(0.01) & 87  & 14 & -11310 & 0.2  & N   & NGC 7448      & 2194 & Sbc                        & 4  \\
1998V  & CN & 1.012(0.007) & -19.31(0.01) & -19.20(0.00) & -19.25(0.01) & 84  & 19 & -10843 & 0.3  & N   & NGC 6627      & 5268 & Sb                         & 3  \\
1998ab & SS & 0.958(0.011) & -19.35(0.02) & -19.24(0.01) & -19.36(0.01) & -   & -  & -9209  & 6.7  & 91T & NGC 4704      & 8134 & Sc                         & 5  \\
1998bu & CN & 1.003(0.005) & -19.09(0.01) & -19.36(0.00) & -19.54(0.01) & 93  & 21 & -11009 & -1.6 & N   & NGC 3368      & 897  & Sab                        & 2  \\
1998dh & BL & 0.937(0.005) & -18.71(0.01) & -18.74(0.01) & -18.82(0.01) & 120 & 24 & -12091 & -0.6 & N   & NGC 7541      & 2678 & Sbc                        & 4  \\
1998ec & BL & 1.029(0.023) & -18.52(0.04) & -18.67(0.05) & -18.66(0.02) & 124 & 11 & -12751 & -3.0 & HV  & UGC 3576      & 5966 & Sb                         & 3  \\
1998eg & CN & 0.970(0.031) & -18.93(0.02) & -18.87(0.01) & -18.94(0.01) & 95  & 20 & -10860 & -0.8 & N   & UGC 12133     & 7423 & Sc                         & 5  \\
1998es & SS & 1.132(0.016) & -19.21(0.01) & -19.26(0.01) & -19.28(0.01) & 51  & 7  & -10183 & -0.5 & 91T & NGC 632       & 3168 & S0                         & 0  \\
1999aa & SS & 1.143(0.026) & -19.33(0.02) & -19.22(0.03) & -19.20(0.01) & 63  & 14 & -10430 & -0.2 & 91T & NGC 2595      & 4330 & Sc                         & 5  \\
1999ac & SS & 1.015(0.010) & -19.01(0.01) & -18.98(0.00) & -19.03(0.00) & 84  & 9  & -9993  & 0.4  & 91T & NGC 6063      & 2848 & Scd                        & 6  \\
1999cc & BL & 0.850(0.012) & -18.88(0.01) & -18.80(0.01) & -18.90(0.01) & 121 & 26 & -12047 & -0.2 & N   & NGC 6038      & 9392 & Sc                         & 5  \\
1999cl & BL & 0.972(0.019) & -17.92(0.02) & -19.03(0.01) & -19.57(0.02) & 133 & 19 & -12395 & 0.7  & HV  & NGC 4501      & 2281 & Sb                         & 3  \\
1999dq & SS & 1.094(0.004) & -19.43(0.00) & -19.45(0.00) & -19.49(0.01) & 51  & 10 & -10680 & 0.4  & 91T & NGC 976       & 4295 & Sc                         & 5  \\
1999ee & SS & 1.124(0.016) & -18.37(0.02) & -18.60(0.08) & -18.71(0.03) & 77  & 9  & -10018 & -0.2 & N   & IC 5179       & 3422 & SA(rs)bc                   & 4  \\
1999ej & BL & 0.792(0.021) & -18.33(0.03) & -18.26(0.01) & -18.32(0.02) & 108 & 27 & -10598 & 1.1  & N   & NGC 495       & 4114 & S0/Sa                      & 0  \\
1999gd & BL & 0.942(0.010) & -17.64(0.02) & -17.97(0.01) & -18.23(0.02) & 108 & 17 & -10713 & 1.1  & N   & NGC 2623      & 5535 & Sa                         & 1  \\
1999gh & BL & 0.756(0.016) & -18.62(0.02) & -19.12(0.03) & -19.07(0.01) & -   & -  & -11467 & 5.5  & N   & NGC 2986      & 2302 & E                          & -5 \\
1999gp & SS & 1.204(0.008) & -19.22(0.01) & -19.19(0.00) & -19.21(0.01) & 52  & 8  & -11332 & 0.0  & 91T & UGC 1993      & 8018 & Sb                         & 3  \\
2000B  & BL & 0.854(0.029) & -18.90(0.03) & -19.14(0.06) & -19.00(0.01) & -   & -  & -      & -    & -   & NGC 2320      & 5901 & E                          & -5 \\
2000E  & SS & 1.101(0.003) & -18.52(0.00) & -18.62(0.00) & -18.66(0.00) & 78  & 13 & -10979 & -2.3 & N   & NGC 6951      & 1424 & SAB(rs)bc                  & 4  \\
2000cn & CL & 0.761(0.009) & -18.44(0.01) & -18.53(0.01) & -18.62(0.01) & -   & -  & -      & -    & -   & UGC 11064     & 7043 & Scd                        & 6  \\
2000cx & SS & 0.863(0.055) & -19.34(0.06) & -19.41(0.10) & -19.29(0.06) & 49  & 3  & -11844 & -0.3 & *   & NGC 524       & 2379 & S0                         & 0  \\
2000dk & CL & 0.762(0.008) & -18.87(0.01) & -18.83(0.01) & -18.90(0.01) & 124 & 47 & -11035 & 0.8  & N   & NGC 382       & 5228 & E                          & -5 \\
2001V  & SS & 1.174(0.021) & -19.56(0.02) & -19.58(0.03) & -19.57(0.02) & -   & -  & -11449 & -3.3 & 91T & NGC 3987      & 4361 & Sb                         & 3  \\
2001el & CN & 0.953(0.012) & -18.14(0.01) & -18.20(0.01) & -18.38(0.01) & 95  & 14 & -11544 & 1.2  & N   & NGC 1448      & 1164 & SAcd                       & 6  \\
2002bo & BL & 0.951(0.024) & -17.92(0.02) & -18.25(0.01) & -18.51(0.04) & 147 & 13 & -13198 & -0.6 & HV  & NGC 3190      & 1271 & SA(s)a pec                 & 1  \\
2004S  & CN & 0.946(0.016) & -18.93(0.02) & -18.78(0.03) & -18.88(0.02) & 89  & 13 & -9256  & 1.6  & N   & MCG -05-16-21 & 2516 & S                      & -2 \\
2006X\footnotemark[$\dag\dag\dag$]  & BL & 0.858(0.025) & -15.51(0.14) & -16.87(0.14) & -17.41(0.14) & 179  & 9 & -15680  & 1.3  & HV & NGC 4321 & 1557 & Sbc                      & 4 \\
\hline
\end{tabularx}
}
\begin{tabnote}
 \footnotemark[$*$] \citet{Branch2009}.\ \footnotemark[$\dag,\ddag,\S,\|$] \citet{Takanashi2008}. The 1 $\sigma$ error is given in between parenthesis. \ \footnotemark[$\#,**$] \citet{Branch2009}.\ \footnotemark[$\dag\dag,\ddag\ddag, \S\S$] \citet{Blondin2012}. \footnotemark[$\dag\dag$] It is the phase of $v_{6355}$ measurement relative to $B-$band maximum. \footnotemark[$\|\|$] \citet{Branch2009} and Transient Name Server <https://wis-tns.weizmann.ac.il> \footnotemark[$\#\#,***$] NASA/IPAC Extragalactic Database <https://ned.ipac.caltech.edu> We obtain $cz$ of SN 2001el, 2001V and 2004S from individual publications (\cite{Krisciunas2003};\cite{Driel2001};\cite{Misra2005}, respectively). \footnotemark[$\dag\dag\dag$] We add the SN 2006X in the list for comparison (see section 4.2). We convert the $\Delta m_{15}(B)$ parameter ($\Delta m_{15}(B) = 1.31\pm 0.05$ mag from \cite{Wang2008}) into $s_{(B)}$ by using the formula $s_{(B)} = (3.06 - \Delta m_{15}(B)) / 2.04$ of \citet{Jha2006}. We calculate the $B, V, R-$band absolute magnitudes using the apparent magnitudes at maximum (\cite{Wang2008}) and the Cepheid distance modulus ($\mu = 30.91\pm 0.14$ mag; \cite{Freedman2001}).
\end{tabnote}
\end{table}
\end{landscape}

\begin{table}[H]
  \captionsetup{font=large}
  \tbl{$\Delta M_B$ and $SNCE$ of the Branch sample}{%
  \label{table:delta_M_B_and_SNCE}
  \centering
  \large
  \begin{tabular}{lcrr}
    \hline
    SN name & Branch sub-type & $\Delta M_B$\footnotemark[$*$] & $SNCE$\footnotemark[$\dag$] \\ 
    \hline
        1989B  & CL & 0.14(0.90)  & 0.38(0.04)  \\
        1991T  & SS & -1.57(0.38) & 0.26(0.04)  \\
        1994D  & CN & 0.20(1.06)  & -0.03(0.02) \\
        1994ae & CN & 0.45(0.51)  & 0.12(0.02)  \\
        1996X  & CN & -0.80(0.31) & 0.03(0.01)  \\
        1997br & SS & -0.66(0.32) & 0.02(0.03)  \\
        1997do & BL & 0.07(0.22)  & 0.04(0.02)  \\
        1997dt & CN & 2.31(0.30)  & 0.51(0.03)  \\
        1998V  & CN & -0.27(0.12) & -0.01(0.01) \\
        1998ab & SS & -0.40(0.09) & -0.02(0.02) \\
        1998bu & CN & -0.06(0.73) & 0.37(0.01)  \\
        1998dh & BL & 0.20(0.24)  & 0.12(0.01)  \\
        1998ec & BL & 0.55(0.12)  & 0.25(0.06)  \\
        1998eg & CN & 0.04(0.11)  & 0.04(0.02)  \\
        1998es & SS & 0.01(0.21)  & 0.16(0.01)  \\
        1999aa & SS & -0.10(0.16) & 0.01(0.04)  \\
        1999ac & SS & 0.04(0.23)  & 0.07(0.01)  \\
        1999cc & BL & -0.16(0.08) & -0.00(0.01) \\
        1999cl & BL & 1.05(0.29)  & 1.21(0.02)  \\
        1999dq & SS & -0.26(0.15) & 0.13(0.00)  \\
        1999ee & SS & 0.84(0.19)  & 0.34(0.08)  \\
        1999ej & BL & 0.24(0.17)  & -0.00(0.03) \\
        1999gd & BL & 1.27(0.12)  & 0.42(0.02)  \\
        1999gh & BL & -0.15(0.29) & 0.56(0.04)  \\
        1999gp & SS & 0.09(0.08)  & 0.09(0.01)  \\
        2000B  & BL & -0.17(0.13) & 0.32(0.07)  \\
        2000E  & SS & 0.66(0.46)  & 0.21(0.00)  \\
        2000cn & CL & 0.04(0.10)  & 0.15(0.01)  \\
        2000cx & SS & -0.59(0.31) & 0.15(0.12)  \\
        2000dk & CL & -0.38(0.13) & 0.02(0.01)  \\
        2001V  & SS & -0.29(0.15) & 0.14(0.04)  \\
        2001el & CN & 0.80(0.56)  & 0.15(0.01)  \\
        2002bo & BL & 1.01(0.51)  & 0.42(0.02)  \\
        2004S  & CN & -0.01(0.24) & -0.06(0.04) \\
        2006X\footnotemark[$\ddag$]  & BL & 3.23(0.15) & 1.44(0.04) \\ \hline
\end{tabular}}
\begin{tabnote}\footnotesize
 \footnotemark[$*$, $\dag$] The 1$\sigma$ error is given in between parentheses.\\
 \footnotemark[$\ddag$] We also include SN 2006X in the list for comparison (see section 4.2).\\
\end{tabnote}
\end{table}
\newpage

\begin{longtable}{lrrr}
\caption{TAK08 photometric sub-sample} \\
\hline 
SN name   & $cz$ (km/s)    & $\Delta M_B$         & $SNCE$  \\
\hline
\endhead
%
1990O  & 9113  & -0.22(0.09) & 0.00(0.03)  \\
1990T  & 11821 & 0.21(0.11)  & 0.25(0.09)  \\
1990Y  & 10664 & 0.85(0.12)  & 0.49(0.11)  \\
1990af & 14837 & -0.35(0.06) & 0.03(0.01)  \\
1991S  & 16213 & -0.39(0.19) & -0.21(0.14) \\
1991U  & 9320  & -0.23(0.10) & -0.02(0.08) \\
1991ag & 4296  & -0.21(0.15) & 0.08(0.04)  \\
1992J  & 13276 & -0.04(0.13) & 0.27(0.22)  \\
1992K  & 3313  & -0.56(0.27) & 0.23(0.24)  \\
1992P  & 7853  & -0.14(0.09) & -0.01(0.02) \\
1992ae & 21966 & -0.16(0.08) & 0.03(0.06)  \\
1992al & 4180  & -0.43(0.16) & -0.04(0.01) \\
1992aq & 29382 & -0.26(0.09) & -0.05(0.04) \\
1992au & 18186 & -0.43(0.26) & -0.00(0.28) \\
1992bc & 5959  & -0.45(0.12) & -0.02(0.03) \\
1992bg & 10381 & -0.20(0.09) & -0.01(0.04) \\
1992bh & 13276 & 0.22(0.06)  & 0.08(0.03)  \\
1992bk & 17021 & -0.31(0.10) & 0.01(0.06)  \\
1992bl & 12697 & -0.42(0.08) & -0.02(0.04) \\
1992bo & 5517  & -0.22(0.12) & 0.01(0.03)  \\
1992bp & 23047 & -0.72(0.08) & -0.07(0.02) \\
1992br & 25696 & -0.52(0.12) & -0.07(0.08) \\
1992bs & 18510 & 0.07(0.06)  & -0.02(0.05) \\
1993B  & 20211 & -0.14(0.06) & -0.05(0.03) \\
1993H  & 7208  & 0.08(0.10)  & 0.20(0.03)  \\
1993O  & 15305 & -0.20(0.06) & -0.04(0.03) \\
1993ac & 14511 & -0.06(0.08) & 0.06(0.08)  \\
1993ae & 5669  & -0.51(0.13) & 0.11(0.05)  \\
1993ag & 14772 & 0.11(0.09)  & 0.10(0.03)  \\
1993ah & 8712  & -0.47(0.14) & -0.27(0.11) \\
1994M  & 6889  & 0.02(0.10)  & 0.03(0.03)  \\
1994Q  & 8595  & 0.07(0.13)  & 0.12(0.10)  \\
1994S  & 4792  & -0.23(0.14) & 0.07(0.03)  \\
1995D  & 2277  & -0.20(0.29) & -0.00(0.02) \\
1995E  & 3499  & 2.23(0.19)  & 0.75(0.03)  \\
1995ac & 14772 & -0.45(0.05) & 0.05(0.01)  \\
1995ak & 6765  & -0.17(0.19) & 0.07(0.08)  \\
1995bd & 4770  & 0.34(0.14)  & 0.39(0.01)  \\
1996C  & 8033  & 0.46(0.09)  & 0.13(0.04)  \\
1996Z  & 2577  & 0.42(0.26)  & 0.34(0.04)  \\
1996ab & 17633 & 1.52(0.08)  & 0.16(0.04)  \\
1996bl & 10664 & -0.24(0.07) & 0.04(0.03)  \\
1996bv & 4970  & 0.14(0.13)  & 0.13(0.02)  \\
1997E  & 3994  & -0.01(0.16) & 0.09(0.01)  \\
1997Y  & 4947  & -0.18(0.14) & -0.04(0.02) \\
1997bp & 2811  & -0.11(0.23) & 0.19(0.01)  \\
1997bq & 2875  & -0.18(0.26) & -0.11(0.11) \\
1997cw & 5272  & 0.30(0.13)  & 0.10(0.04)  \\
1997dg & 10060 & 0.02(0.09)  & -0.03(0.02) \\
1998dk & 3597  & 0.21(0.20)  & 0.16(0.08)  \\
1998dx & 14772 & -0.36(0.09) & -0.07(0.04) \\
1998ef & 4992  & -0.59(0.13) & 0.02(0.02)  \\
1999X  & 7439  & -0.14(0.22) & -0.07(0.17) \\
1999cp & 3079  & -0.15(0.21) & -0.00(0.01) \\
1999dk & 4161  & 0.09(0.16)  & 0.08(0.01)  \\
1999ef & 11153 & 0.08(0.07)  & -0.04(0.04) \\
1999ek & 5248  & 0.05(0.13)  & 0.12(0.01)  \\
2000bh & 6796  & -0.01(0.10) & 0.05(0.03)  \\
2000bk & 7541  & 0.28(0.09)  & 0.28(0.01)  \\
2000ca & 7015  & -0.26(0.09) & -0.02(0.01) \\
2000ce & 4858  & 1.82(0.16)  & 0.50(0.07)  \\
2000cf & 10664 & -0.14(0.08) & -0.01(0.03) \\
2001ba & 9154  & -0.32(0.07) & -0.03(0.01) \\
2001bt & 4355  & 0.12(0.15)  & 0.20(0.01)  \\
2001cn & 4621  & 0.05(0.14)  & 0.15(0.01)  \\
2001cz & 4902  & -0.07(0.13) & 0.15(0.01)  \\
2001en & 4476  & 0.15(0.15)  & 0.16(0.01)  \\ \hline
\end{longtable}

\clearpage
\section{ Discussions }
\subsection{ Color-color diagram }
First, we discuss the $B-V$, $V-R$ color-color diagram.
In figure \ref{fig:cc_host_stretch_plot}, the distributions with host morphology show that SNe Ia that occur in early-type spirals, rather than late-type spirals, tend to have redder colors along with the dust extinction lines. The distribution trend mentioned above holds when divided into Branch sub-types (see figure \ref{fig:cc_host_plot}). 
In  figure 1 of  \citet{Childress2013}, using 115 SNe Ia from the Nearby Supernova Factory, they compared SNe Ia color, host galaxy stellar mass, specific star formation rate (sSFR) and metallicity. They found that SNe Ia with red color belong to galaxies with intermediate stellar mass and sSFR. Our findings are consistent with their result. In the Branch CN sample, there is no outlier object which deviates from the extinction lines. On the other hand, there are three outliers in BL sample (SN 1999cl, 1999gh and 2000B) and the two of them (SN 1999gh and 2000B) occur in elliptical galaxies.
This implies that we can roughly distinguish CN and peculiar red BL SNe Ia with a color-color diagram. \citet{Mandel2014} reported that BL high-velocity (=HV) SNe Ia show intrinsic redder $B-V$ color (by $\sim 0.06$ mag) than normal-velocity events. In addition, smaller $R_V$ are predominantly derived from HV SNe Ia (\cite{Wang2009}) and some of our BLs in figure \ref{fig:cc_plot} are consistent with their results.
\subsection{ Outliers on the color-color and $\Delta M_B - SNCE$ diagrams }
We discuss some outliers on the color-color and $\Delta M_B - SNCE$ diagrams. First we focus on the BL sub-type outliers: SN 1999cl, 1999gh and 2000B. The Si {\scriptsize II} $\lambda 6355$ velocity of SN 1999gh is $v_{6355}$ = $-11,467$ km/s, which is relatively small among BL sub-type (SN 2000B has no measurements). The boundary between NV and HV SNe Ia is located at $\sim -12,200$\ km/s (see figure 8 of \cite{Blondin2012}). There is a large overlap between the CN and the NV SNe Ia, as well as between the BL and HV SNe Ia. So the Si {\scriptsize II} velocity may not be a major quantity to discriminate peculiar BL objects.
SN 1999gh and 2000B have red color in $B-V$  but bluer in $V-R$ than SS objects. The host morphological types of the SN 1999gh and 2000B are both elliptical galaxies ($T = 5$), so they are in generally less-dusty environment. Therefore, it can be inferred that SN 1999gh and 2000B have intrinsic red $B-V$ colors considering their blue $V-R$ colors. \citet{Mandel2014} shows similar result that intrinsic color distributions of HV and normal SNe Ia exhibit significant discrepancies in $B-V$ and $B-R$ colors.\par
SN 1999cl has distinct red color ($B-V = 1.11, V-R = 0.54$) compared with the other BL objects. Its host galaxy type is Sb ($T = 3$), which often shows strong extinction. From the image of its discovery, SN 1999cl is located in the disk component of its host. The Si {\scriptsize II} $\lambda 6355$ velocity of SN 1999cl is $v_{6355}$ = -12,395 km/s, which is typical among BL sub-type, but the overall shape of its spectrum is much redder than others. Based on these facts, SN 1999cl can be suffered from significant host extinction. In \citet{Krisciunas2006}, they obtained the value of $R_V = 1.55$ for SN 1999cl, which may imply the size of dust is very small ($\sim 10^{-3} \mu m$). But when we compare SN 1999cl with SN 1999gh and 2000B on the $\Delta M_B - SNCE$ diagram, another possibility is that SN 1999cl has intrinsically red color, and also is strongly reddened by Milky-Way type dust extinction. Given the distribution trend of sample with Sa-Sb host galaxies seen in figure \ref{fig:dm_snce_host}, the latter is more likely.\par
We compared SN 1999cl with SN 2006X, another high-extinction SN Ia (\cite{Wang2008}) and also is classified as BL (EW(6100), EW(5750) = 179, 9) in \citet{Branch2009}. The host of SN 2006X (= NGC4321) is an Sbc-type galaxy, which is similar to that of SN 1999cl. In \citet{Blondin2009}, they report that the variability in Na {\scriptsize I} D lines was found both in the spectra of SN 1999cl and 2006X. The Na {\scriptsize I} D variability is interpreted to be associated with CSM in the SN Ia progenitor system (\cite{Patat2007}). SN 2006X has $B-V = 1.35$ and $V-R = 0.56$ colors at $B-$band maximum (\cite{Wang2008}). 
The $V - R$ color of SN 2006X is nearly the same as SN 1999cl but the $B - V$ color is $\sim 0.2$ mag redder than that of SN 1999cl. The $\Delta M_B$ and $SNCE$ of SN 2006X are $\Delta M_B = 3.23\pm 0.15$ and $SNCE = 1.44\pm 0.04$\footnote{ We used the Cepheid-based distance modulus of $30.91\pm 0.14$ mag (\cite{Freedman2001}) for calculating $\Delta M_B$.} (see figure \ref{fig:dm_snce_host} and \ref{fig: dmb_snce_all_sample}). The slope between SN 1999cl and 2006X is close to the slope of $R_V = 3.1$ line on the $\Delta M_B - SNCE$ diagram, suggesting that both SN 1999cl and 2006X have intrinsically red color (possibly by CSM) and they are highly extincted by interstellar dust.
\par
While most SS objects show bluer color in both $B-V$ and $V-R$, SN 1997br has bluer color in $B-V$ but redder color in $V-R$ (the fourth reddest color in the Branch sample in figure \ref{fig:cc_plot}). Its host galaxy type is Sd/Irr ($T = 7$) and the morphological index is the largest in our sample, suggesting that the SN 1997br is in dusty environment. However, if considering its bluer color in $B-V$, the peculiar color may be intrinsic. In \citet{Li1999}, they reported that the light curve and time evolution of the spectra of SN 1997br are similar to those of over-luminous type of SN 1991T. However, their locations in color-color diagram are different.\par
SN 1997dt suffers highest extinction but has a typical value of stretch factor ($s_{(B)} = 0.971$) among CN sample. In addition, the spectral features of SN 1997dt look similar to the other CNs but the flux of the spectrum was reduced in shorter wavelengths. Its host galaxy type is an early-spiral Sbc ($T = 4$) which often shows strong extinction.
\subsection{ Implications for intrinsic color of SNe Ia }
Since $\Delta M_B$ and $SNCE$ are residuals from the stretch-magnitude/color relations of the bluest SNe Ia, $\Delta M_B - SNCE$ diagram tells us the information of intrinsic luminosity/color and extinction by dust. It is clear that there are three BL objects (SN 1999cl, 1999gh and 2000B) which have large $SNCE$ in figure \ref{fig:dmb_snce_Branch_sample}. This means they show red color after correcting for $s_{(B)}$. As shown in figure \ref{fig:cc_host_stretch_plot}, these three BLs have small stretch factors in the Branch sample. When stretch correction is performed, their colors move in the lower-left direction in figure \ref{fig:cc_host_stretch_plot} (shown as the blue arrow). In addition, two-thirds of the BLs (SN 1999gh and 2000B) occur in early-type (E/S0) galaxies, suggesting that BL objects whose hosts are E/S0 galaxies need to apply different stretch correction. Another possibility is that the progenitor scenarios (explosion mechanisms) of such BL objects may differ from typical SNe Ia.\par
It is possible that small $R_V$ can be derived because the $R_V$ is averaged both for normal SNe Ia and intrinsically red SNe Ia. We employ the Bayesian linear regression method by \citet{Kelly2007} with its python package \textit{linmix\footnote{https://github.com/jmeyers314/linmix}}. In figure \ref{fig:dmb_snce_Branch_sample}, the solid blue line shows the average of \textit{linmix} MCMC fit results to the Branch sub-sample excluding three BL outliers (SN 1999cl,1999gh and 2000B) with 100 random MCMC fit lines overlaid as light blue. As a result, the average regression line which corresponds to $R_V = 3.3^{+0.8}_{-0.6}$ well matches the Milky-Way like extinction with $R_V = 3.1$.\par
To make the result more robust, we include the TAK08 photometric sub-sample for the regression analysis. In figure \ref{fig: dmb_snce_all_sample}, the green solid line shows the average of 100 MCMC fit results using all Branch sub-sample and TAK08 photometric sub-sample while the blue line shows the same result without three BL outliers (SN 1999cl, 1999gh and 2000B). The slope of the green line corresponds to $R_V = 2.3 \pm 0.3$, while that of the blue line is $R_V = 3.0^{+0.4}_{-0.3}$ which is almost identical to $R_V = 3.1$. Therefore the three BL outliers make $R_V$ small. When we exclude such outliers, the extinction law for SNe Ia becomes close to $R_V = 3.1$.\par

When we use SNe Ia for cosmological studies to measure distance, excluding peculiar BL objects may give more accurate results. This could be done just to use two optical colors (e.g., $B-V$ and $V-R$) around the time of maximum brightness. In addition, further environmental studies will give us clues about the difference in progenitor scenarios among Branch sub-types.\par
Using early phase color information, \citet{Stritzinger2018} found that there are two distinct early populations with different early color evolution in $B-V$ and all blue events are of the Branch SS sub-type, while all early red events except for the peculiar 2000cx-like SN 2012fr are of the Branch CN or CL sub-types. It is also important to study the early color evolution with Branch sub-types.
\section{ Summary }
In this study, with Branch spectroscopic classification as well as host galaxy morphology, we investigate the diversity of color and dust extinction of nearby 34 SNe Ia ($z \lesssim 0.04$). We summarize our findings as below.
\begin{itemize}
    \item In the $B-V, V-R$ color-color diagram, different distribution among different host galaxy morphology can be seen: SNe Ia which occur in early-type spirals have the reddest color. The distribution trend holds when divided into the Branch spectroscopic sub-types.
    \item The three BLs (SN 1999cl, 1999gh and 2000B) have red colors in the $\Delta M_B - SNCE$ diagram. Two of them (SN 1999gh and 2000B) can be explained by their intrinsic red colors, and their hosts are elliptical galaxies which usually have little interstellar dusts. On the other hand, 1999cl which occur in Sb type host can be explained by both its intrinsic red color and strong extinction by interstellar dust.
    
    \item In our sample, it is inferred that lower value of $R_V$ parameter is not necessary to explain the color diversity of SNe Ia. In other words, it suggests that the extinction law for most of SNe Ia might be explained by the typical extinction law in the Milky Way ($R_V = 3.1$).
\end{itemize}
Our results suggest the possibility of typical host galaxy extinction law for SNe Ia as seen in the Milky Way. It was suggested by TAK17 that there seems to be two (or more) different sub-groups with different intrinsic colors and/or dust extinction laws, but we infer this may be caused by some objects with peculiar intrinsic red color.
\section{ Acknowledgments }
We thank the anonymous referee for reading the paper carefully and providing thoughtful comments, many of which have resulted in changes to improve the revised version of the manuscript. This work was supported by JSPS KAKENHI Grant numbers 18H05223 and 18H04342. N.A. gratefully appreciates the financial support of Hattori International Scholarship Foundation (HISF) for a grant that made it possible to complete this study.
\bibliographystyle{apj}
\begin{bibliography}{pasj}
\end{bibliography}
\end{document}